\documentclass[journal]{IEEEtran}
\usepackage{graphicx}
\usepackage{cite}
\usepackage{bm}
\usepackage{booktabs}
\usepackage{amsmath}
\usepackage{epsfig}
\usepackage{color,soul}
\usepackage{url}

\usepackage{breakurl}

\ifCLASSINFOpdf
\else
\fi
\begin{document}
%
\title{Detecting Medical Misinformation on Social Media Using Multimodal Deep Learning}
%
%
%

\author{Zuhui~Wang, Zhaozheng~Yin,~\IEEEmembership{Member,~IEEE,} and Young~Anna~Argyris
\thanks{Zuhui~Wang is with the Department of Computer Science, Stony Brook University, Stony Brook, NY 11794 USA. (e-mail: zuwang@cs.stonybrook.edu)}
\thanks{Zhaozheng~Yin is with the AI Institute,  Department of Computer Science, Department of Biomedical Informatics, and Department of Applied Mathematics \& Statistics (Affiliated), Stony Brook University, Stony Brook, NY 11794 USA. (e-mail: zyin@cs.stonybrook.edu)}
\thanks{Young~Anna~Argyris is with the Department of Media \& Information, Michigan State University, East Lansing, Michigan, 48824 MI USA. (e-mail: argyris@msu.edu)}
}
\maketitle

\begin{abstract}
In 2019, outbreaks of vaccine-preventable diseases reached the highest number in the US since 1992. Medical misinformation, such as antivaccine content propagating through social media, is associated with increases in vaccine delay and refusal. Our overall goal is to develop an automatic detector for antivaccine messages to counteract the negative impact that antivaccine messages have on the public health. Very few extant detection systems have considered multimodality of social media posts (images, texts, and hashtags), and instead focus on textual components, despite the rapid growth of photo-sharing applications (e.g., Instagram). As a result, existing systems are not sufficient for detecting antivaccine messages with heavy visual components (e.g., images) posted on these newer platforms. To solve this problem, we propose a deep learning network that leverages both visual and textual information. A new semantic- and task-level attention mechanism was created to help our model to focus on the essential contents of a post that signal antivaccine messages. The proposed model, which consists of three branches, can generate comprehensive fused features for predictions. Moreover, an ensemble method is proposed to further improve the final prediction accuracy. To evaluate the proposed model's performance, a real-world social media dataset that consists of more than 30,000 samples was collected from Instagram between January 2016 and October 2019. Our 30 experiment results demonstrate that the final network achieves above 97\% testing accuracy and outperforms other relevant models, demonstrating that it can detect a large amount of antivaccine messages posted daily. The implementation code is available at \fontfamily{cmtt}\selectfont\textcolor{magenta}{https://github.com/wzhings/antivaccine\_detection}.
\end{abstract}

\begin{IEEEkeywords}
Antivaccine detection, attention mechanism, multimodal feature fusion, ensemble method 
\end{IEEEkeywords}

%
\IEEEpeerreviewmaketitle

\section{Introduction}
%
%
%
%

\IEEEPARstart{C}{hildhood} vaccine hesitancy is a complex public health problem in the United States (US)~\cite{r1}. The increasing number of under-vaccinated children is linked to parental delay or refusal of vaccines. These vaccine-hesitant parents often seek health information (including vaccines) on social media. Social media applications have thus contributed at least in part to the rapid growth of the antivaccine movement~\cite{r37}. Recently, image-sharing social media platforms (such as Instagram, Snapchat, and Pinterest) are growing in popularity. Users of these platforms include young mothers from low-income and underserved populations who are more likely to trust social media for health information~\cite{paige2017}. To counteract the large-scale spread of antivaccine messages to these vulnerable populations, it is urgent to develop an automatic and intelligent antivaccine detection system.

We chose Instagram as our target image-sharing social media platform because Instagram is leading among all other image-sharing platforms with 500 million daily active users~\cite{Perrin2019}. The majority of Instagram users consist of young women from diverse backgrounds~\cite{paige2017}, thus suitable for our purpose of counteracting the propagation of antivaccine posts via social media.

\begin{figure}[!t]
\centering
\includegraphics[width=\linewidth]{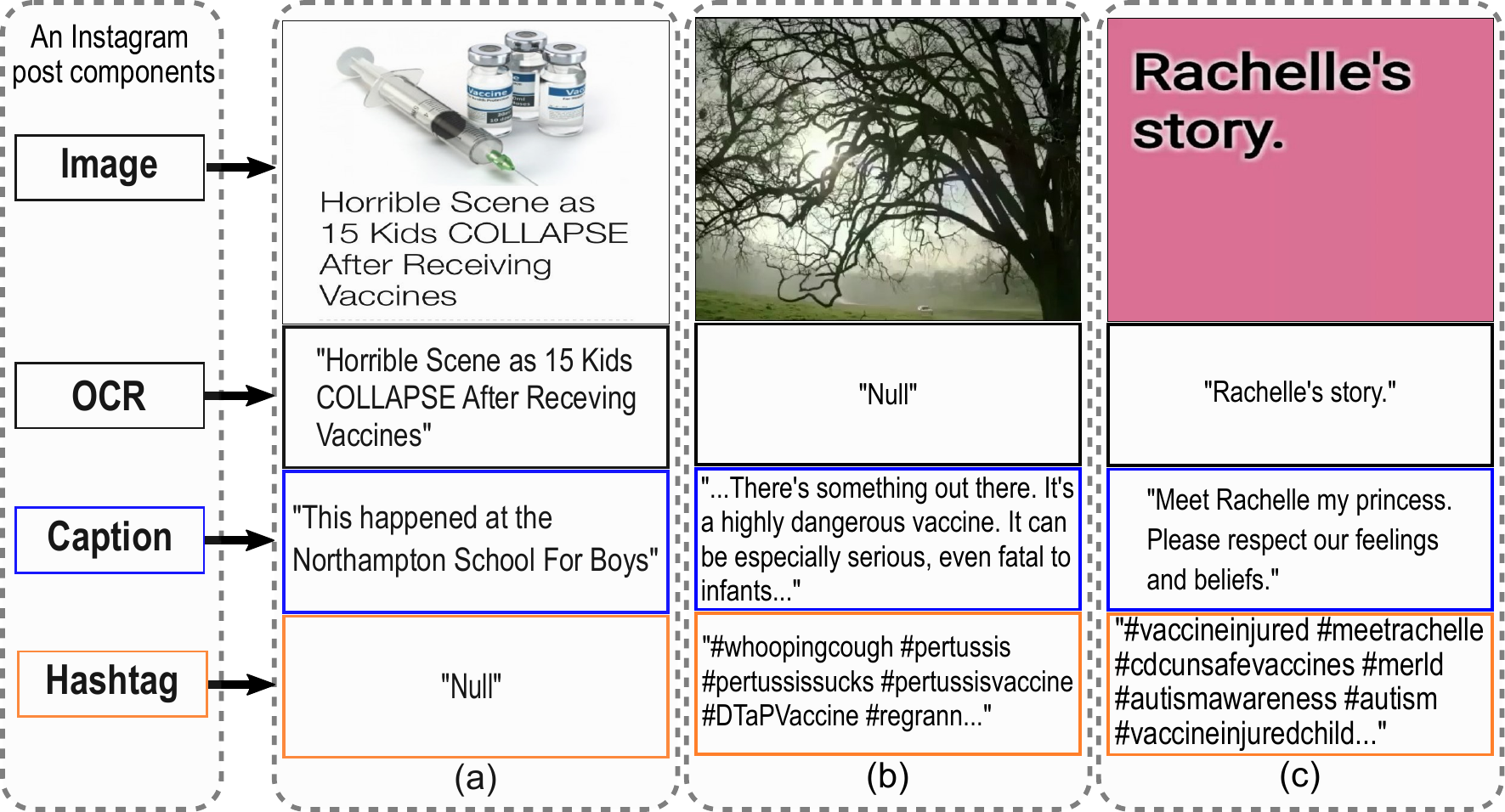}
\caption{Examples of three antivaccine posts on Instagram. Instagram posts contain images (photographs and texts overlaid in images), captions (i.e., a poster's comments), and hashtags. In~(a), if only the caption and hashtags are considered, excluding the image (e.g., the syringe and the texts embedded in the image), this post cannot be detected as an antivaccine post. In~(b), if only the image that contains a tree is considered, but the caption (e.g., ``a highly dangerous vaccine'') is not, this post cannot be predicted as an antivaccine post. In~(c), if the image, text in the image, and the caption were considered, but the hashtags (e.g., \#cdcunsafevaccine and \#vaccineinjuriedchild) are not, this antivaccine post cannot be detected as antivaccine. (``Null'' denotes that no textual content is detected in the post, and ``OCR'' denotes Optical Character Recognition.)}
\label{fig:r1}
\end{figure}

Instagram, like other image-sharing platforms, allows users to include multimodal content in their posts, including images, captions, and hashtags, as illustrated in Fig.~\ref{fig:r1}. These multimodal elements in an Instagram post largely encompass (1)~visual elements (e.g.,~photographs and posters), and (2)~textual elements (captions, words in images, and hashtags). All of these are important for detecting antivaccine themes in a social media post. First, images draw more engagement from the audiences (e.g.,~likes and comments) than texts alone~\cite{Eftekhar14}. Captions in posts are unarguably an important means through which the users deliver the main theme of their posts. However, given the increasing surveillance for misinformation on social media, savvy users opt for overlaying texts on an image to subvert the regulation~\cite{Chancellor16,Laat2017}. Lastly, many social media users employ hashtags like catchphrases wherein they succinctly summarize their main claims in their attempt to make their posts identifiable by like-minded crowds~\cite{synnott2017}. Given the importance and relevance of visual and textual elements in antivaccine messages, it is necessary to leverage all the multimodal information in social media posts. The neglect of an element could result in failures in detecting antivaccine messages, as indicated in Fig.~\ref{fig:r1}(a) through Fig.~\ref{fig:r1}(c).

Nonetheless, most prior studies only involve a small subset of textual components (i.e., posters' comments) when investigating antivaccine messages~\cite{MitraICWSM2016,BjarkeCoRR2019,r7}. The lack of comprehensive multimodality consideration in these prior studies therefore limits their abilities to accurately capture the antivaccine theme in social media posts, which are increasingly visual and textural. 

The primary goal of this study, therefore, is to develop a reliable multimodal network to detect antivaccine messages on social media. Considering multimodality to achieve highly accurate predictions is essential for detecting antivaccine messages because both false positives and false negatives can have detrimental effects. Specifically, frequent false positives will lead to backlash from social media users who feel that their posts are unfairly demoted and censored; false negatives will fail to suppress the propagation of antivaccine messages. 

The proposed model architecture entails multiple branches to extract features of images, words in images, captions, and hashtags in social media posts. The multimodal features are fused using attention mechanisms to produce comprehensive representations for the final detection. For the antivaccine detection task, the attention mechanism is implemented as a neural network model layer that helps the model to assign more weights to the essential content of antivaccine messages and improves the proposed model performance. None of the prior studies have considered all of these in one study. To develop and validate our model, we collected more than 30,000 Instagram posts. The resultant detection accuracy rate of our proposed model is the highest reported in the literature.

To summarize, our main contributions are as follows: (1) an end-to-end multimodal feature fusion network; (2) two types of attention mechanisms applied to visual and textual contents; and (3) two modules for projecting and fusing multimodal features. These contributions fill an important gap in the prior studies by considering multimodal information in antivaccine social media posts and help improve the algorithmic detection of medical information.

\section{Related Work}
We review recent works about misinformation detection on social media, such as fake news, hate speeches, health misinformation, and finally antivaccine messages. In the literature review, we focus on the strengths and weaknesses of each approach.

\subsection{Fake News Detection}
The innovative fake news detection models have leveraged both visual and textual information, but still are limited in the following aspects. For example,
\begin{itemize}
\item A recurrent neural network (RNN) was proposed to fuse images, text, and social contexts with attention mechanisms in order to detect fake news~\cite{r26}. A limitation of this approach is that text information and social contexts in the post were represented by the same method. Moreover, the attention mechanism was only applied to visual components, not to the textual counterparts. 
\item To enable proposed models to have greater generalizability, a deep learning network with memory blocks was proposed to save event-invariant information~\cite{ZhangACMMM2019}. For an input message, the model can output features shared among the same events and make decisions on the shared features. However, this method did not have an effective way to process hashtags in the message. 
\item A generative adversarial network was proposed to generate event-invariant multimodal features to detect fake news~\cite{r21}. A text-based convolutional neural network (text-CNN) was proposed to extract textual features in the message. However, textual context information was not considered in this approach. 
\item A multimodal, variational auto-encoder coupled with a binary classifier was proposed to detect fake news~\cite{KhattarWWW2019}. An encoder and a decoder were developed to learn a shared multimodal representation of each message. Nevertheless, hashtag information was not employed during the model training.
\end{itemize}

\subsection{Hate Speech Detection}
Hate speech refers to objectionable speech that encourages violence or expresses hate towards people~\cite{Schmidt2017}. Some prior studies have employed deep learning models that leveraged multimodal information to detect hate speech but with room for improvements, like in fake news detection systems. Some researchers proposed to fuse multimodal information by the simple concatenation. For example, 
\begin{itemize}
    \item To detect misogynistic contents on social media, in~\cite{Gasparini2018}, a pre-trained AlexNet model~\cite{Krizhevsky2012} was proposed to extract visual features from images, and a pre-trained word2vec embedding model~\cite{Mikolov2013} was selected to represent textual features in the posts. The two types of extracted features were then concatenated directly to be processed to predict whether the input post contained misogynistic information or not. The pre-trained models were not considered to be fine-tuned during the model training.
    \item A deep learning model that leveraged both visual and textual information to detect hate speech on Internet memes~\cite{Sabat2019}. A pre-trained VGG16 model~\cite{Simonyan15} was employed to extract visual features from images. The text in images was first extracted by an Optical Character Recognition (OCR) algorithm. The OCR results were then processed by a pre-trained BERT system~\cite{Devlin19}. After that, the extract visual and textual features were concatenated directly to be classified whether the input is a hate speech post or not. The pre-trained models were not fine-tuned during the model training, and no attention mechanism was employed.
\end{itemize}

To fuse multimodal features effectively, some other researchers proposed to capture interactions between features from different modalities. For example, \begin{itemize}
    \item A neural network that leveraged multimodal information was proposed to detect hate speech on social media~\cite{kumari20}. A fine-tuned VGG16 model was employed to extract visual features from images, and a 1-D convolutional neural network to extract features from text contents. These features were then optimized by a genetic algorithm to select the better performing subset of features for further model processing. Despite these strengths, the textual context information was not considered to extract during the model training.
    \item A pre-trained ResNet~\cite{r10} was proposed to extract visual features from images, and a 1-D convolutional layer was used to extract textual information~\cite{yang2019}. Besides, an attention mechanism was proposed to generate better textual context information based on the image features. Then, the features of different modalities were fused together to detect hate speech posts. The weaknesses of this work center on neglecting textual context information in recurrent layers. Also, the text in images was not considered.
    \item The textual tweet contents and text in images were extracted by a recurrent neural network separately, and the visual features were extracted by a fine-tuned InceptionV3 model~\cite{Szegedy16} from images~\cite{Gomez20}. Then, the multimodal features were convolved to generate fused features for the model to predict whether the input data contained hate speech content or not. The shortcoming of this study is that the high-level textual context information was not considered to be extracted by more powerful bidirectional recurrent layers, and no attention mechanism was employed in the proposed model.
\end{itemize}

\subsection{Health Misinformation Detection}
Health misinformation on social media has been a subject of particular concern because users seek medical information from social media as opposed to specialized healthcare websites~\cite{Choudhury14}. Although additional efforts are required to check the source credibility, users may not spend a lot of time and energy verifying the source expertise~\cite{eysenbach08}. Some research works have sought to identify the factors that affect the perceived credibility of online health information~\cite{Viviani17}. Machine learning technologies have been used to identify sources that are credible based on their relationship to professional healthcare sources~\cite{Abbasi13} and their reputation based on followers and shares~\cite{Weitzel14}. For example, 
\begin{itemize}
    \item Machine learning models were developed to track Zika fever misinformation on Twitter~\cite{Ghenai17}. The authors selected the top ten features of Zika fever from a feature set and then built a random decision tree model to classify whether the input tweet is a rumor or not. 
    \item Machine learning models were also proposed to detect health misinformation on Chinese online social media~\cite{Liu19}. For the data samples, 75 features that were collected based on specific word frequencies were used as input features. Then, a gradient boosting decision tree model~\cite{friedman2001} was built to classify whether the input sample is reliable or not. 
\end{itemize}
These methods did not leverage embedding models to vectorize input texts, and the context features among data and hashtag information are overlooked.

\subsection{Antivaccine Detection}
Several prior studies have suggested machine-learning approaches to investigate antivaccine messages. For example, 
\begin{itemize}
    \item A linear regression model was trained to classify antivaccine messages on Twitter~\cite{r7}. After identifying antivaccine tweets, they began to explore the temporal trends, geographic distribution, and demographic correlates of antivaccine attitudes on Twitter.
    \item Three different clustering algorithms (visualization of similarities, Louvain, and {\it k}-means) were used to find and analyze similar antivaccine cases~\cite{botsis2014}. This approach was able to find shared clinical characteristics from similar antivaccine cases.
    \item After collecting participants' attitudes towards the vaccination debate on Twitter, a support vector machine (SVM)~\cite{CortesV95} was employed in~\cite{MitraICWSM2016} to classify Twitter data and concluded that new interventions are needed to correct misleading vaccination related claims.
    \item An RNN was trained to detect and analyze antivaccine sentiment clusters on Twitter~\cite{BjarkeCoRR2019}. They observed profiles of both antivaccine and provaccine accounts. They discovered, for example, that there is a strong link between antivaccine accounts and commercial sites that sell alternative health products. 
\end{itemize}
However, these methods only consider textual information when analyzing vaccine misinformation on social media. Besides, hashtags have not been used in their methods. Moreover, it has been demonstrated that multimodal features can improve predictive performance in various health-related tasks. For instance,
\begin{itemize}
    \item A stacked deep polynomial network that leverages both magnetic resonance imaging (MRI) data and positron emission tomography (PET) data to detect Alzheimer's disease~\cite{Shi2018}.
    \item A convolutional neural network that leverages the movements from patients' speech, handwriting, and gait to evaluate the Parkinson's disease~\cite{Correa2019}.
    \item A deep convolutional neural network that leverages clinical and dermoscopic images and patient metadata to classify the seven-point melanoma checklist criteria and perform skin lesion diagnosis~\cite{Kawahara2019}.
\end{itemize}
These methods demonstrated that the performance of multimodal networks outperform all the single-model networks. Accordingly, to detect antivaccine messages, those methods that only consider textual information would be limited in processing social media messages that are increasingly visual and shorter with only hashtags in lieu of lengthy narratives~\cite{song2018}.

\subsection{Motivations}
In our study, we alleviate each of the limitations found in the prior studies. The pre-trained models for visual and textual branches are fine-tuned during model training. Attention mechanisms are applied to both visual and textual branches to obtain stronger feature representations. Especially, we leverage multimodal information to our detection of antivaccine posts for its proven effectiveness in detecting health information.
\begin{figure}[!t]
\centering
\includegraphics[width=\linewidth]{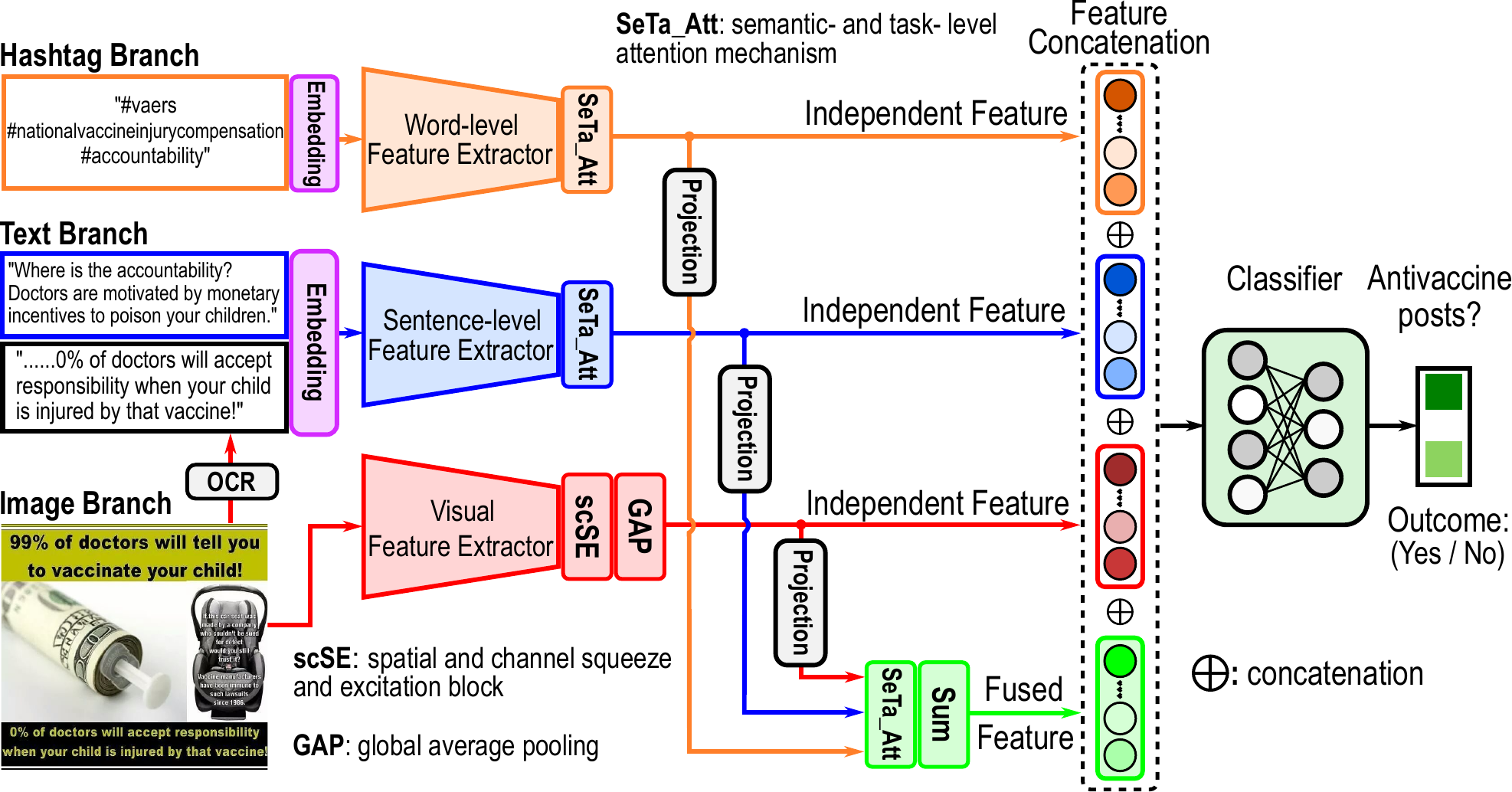}
\caption{The overview architecture of the proposed multimodal network for antivaccine message detection.}
\label{fig:r2}
\end{figure}

\section{Methodology}
\subsection{Model Overview}
A multimodal network was proposed to detect antivaccine messages on social media. As illustrated in Fig.~\ref{fig:r2}, the proposed model contained three main branches for independent feature extraction: the hashtag branch, the text branch, and the image branch. The text overlaying the image was extracted by an optical character recognition (OCR) algorithm and was provided as an additional input to the text branch. Moreover, attention mechanisms were employed separately in each component. The three independent, single-modal features were then projected and fused to generate a fused feature. Finally, the three independent, single-modal features and the fused feature were concatenated for the final classification. In the following, we describe the details of each module in the multimodal network and then present an ensemble of the multimodal networks. 

\subsection{The Hashtag Branch}
Hashtags are used to express the main topic of the post and to spread posts to those interested in the topic~\cite{synnott2017}. Hashtags can be viewed as keywords for Instagram posts. As indicated in Fig.~\ref{fig:r1}, these hashtags generally have broad meanings individually and do not have rich context information as words in natural English sentences. If a post is related to antivaccine messages, it will often contain hashtags linked to antivaccine information, such as \textit{\#vaccinefree}. 
	
Hashtags have no white space between consecutive words. It is difficult to use a regular word embedding model to get an effective hashtag vector representation because hashtags will trigger the out-of-vocabulary issue. To represent hashtags appropriately, in this paper, we employed the fine-tuned fastText word embedding model to represent hashtags~\cite{r28}. This model summarizes subword embeddings as word representations if new words cannot be found directly. For example, \textit{\#vaccineinjury} cannot find any vector representation directly from the embedding vocabulary. However, \textit{vaccine}, \textit{injury} can be found. Therefore, hashtags will be assigned an embedding vector representation after using the fastText model. After obtaining the vector representation of every hashtag in the post, we used a dense layer with the \textit{tanh} activation function to generate hidden representation ${\bm h}^*$ on the post's hashtags:
\begin{equation}
{\bm h}^*_n = \textbf{Dense}({\bm E}_{\textit{fast}}{\bm o}_n), \ n \in [1, N]
\label{conv_hash}
\end{equation}
where $N$ is the number of hashtags in the post. ${\bm o}_n$ is the one-hot encoding representation of the $n$-th hashtag in the caption. ${\bm E}_{\textit{fast}}$ is the fine-tuned embedding matrix. The details of fine-tuning are described in Section~\ref{sec:IC}. The summation of all the hashtag representations was then treated as the representation of hashtags of this post:
\begin{equation}
{\bm F}_H = \sum_{n=1}^{N}{\bm h}^*_n, \ n \in [1, N]
\end{equation}
\begin{figure}[!t]
\centering
\includegraphics[width=0.95\linewidth, height=0.19\textheight]{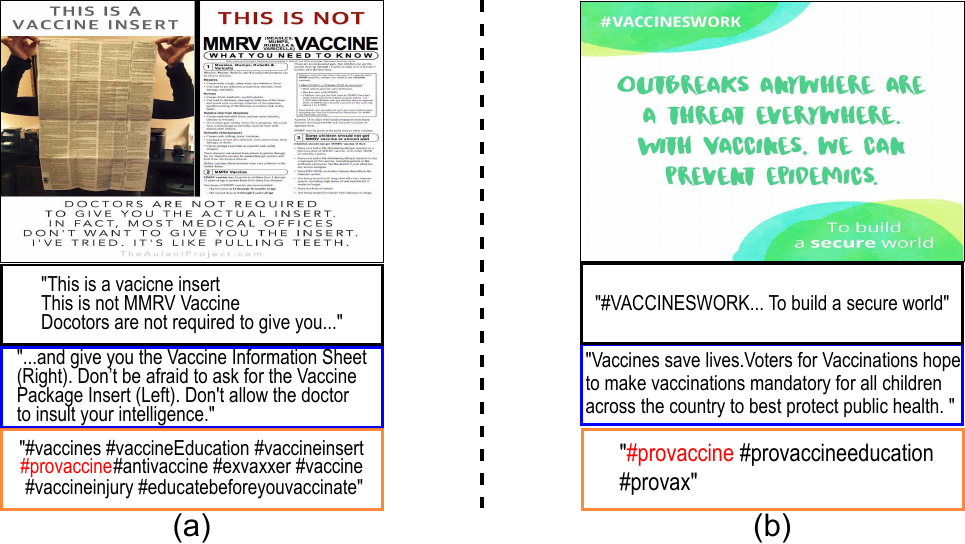}
\caption{Example posts that contain \#provaccine and have antithetical attitudes towards vaccination.}
\label{fig:r3}
\end{figure}

Hashtag information is useful for detecting antivaccine messages, but it is not reliable to conduct antivaccine detection with hashtag information only. There are two straightforward reasons: (1) some hashtags have a neutral attitude such as \textit{\#vaccine}, and both antivaccine and provaccine messages tend to attach this kind of hashtag; (2) antivaccinists may use provaccine hashtags to disturb vaccination supporters or followers. For example, in Fig. \ref{fig:r3}, both posts contain \textit{\#provaccine}; however, these two posts state antithetical attitudes towards vaccination. Thus, vaccine-related hashtags are useful to identify vaccine-related posts from vaccine-unrelated posts, but we require other information from the caption and image of a post to determine if this post is related to antivaccine or provaccine messages.

\subsection{Attention Mechanisms in the Texts (the SeTa\_Att Module)}
\label{sec:33}
\begin{figure}[!t]
\centering
\includegraphics[width=0.8\linewidth]{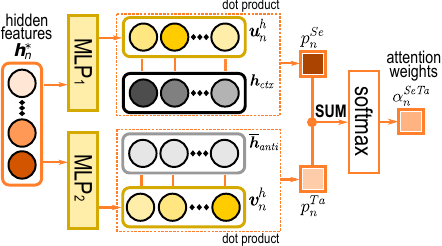}
\caption{The computation procedure of SeTa-attention weight for the hashtag branch.}
\label{fig:r4}
\end{figure}

Attention mechanisms are commonly implemented as neural network model layers. They are used to help the model focus on the important content of the layer input and generate better layer representations as the layer output. To better represent textual contents in the post, we propose a new attention mechanism called semantic- and task-attention (SeTa) that will enable our model to pay attention to discriminative textual contents based on both semantic-level and task-level information. In the hashtag branch, for \textit{semantic}-level attention, the new attention mechanism will indicate which hashtags contribute the most to representing the post’s meaning; in contrast, the \textit{task}-level attention also concerns which hashtags play an essential role in discriminative antivaccine message detection. Thus, the hashtags that contribute the most to expressing the post's meaning and helping detect antivaccine messages deserve the majority of the attention. The procedure to compute the SeTa-attention weight for hashtags is summarized in Fig.~\ref{fig:r4}.

Firstly, to compute the semantic-level attention, we fed the representation of hashtag ${\bm h}^*_n$ into a multilayer perceptron (MLP$_1$) to get its corresponding hidden states ${\bm u}^h_n$:
\begin{equation}
{\bm u}^h_n = tanh({\bm W}_{H_1} {\bm h}^*_n + {\bm b}_{H_1}), \ n\in[1, N]
\end{equation}
where ${\bm W}_{H_1}$ and ${\bm b}_{H_1}$ are the learnable weight matrix and bias term, respectively. We then computed the semantic importance of the hashtag, $p_n^{\textit{Se}}$:
\begin{equation}
p_n^{\textit{Se}} = {{\bm u}^h_n}^T{\bm h}_{\textit{ctx}}, \ n\in[1, N]
\end{equation}
where ${\bm h}_{\textit{ctx}}$ is a learnable context parameter vector.  

The task-level attention for hashtag ${\bm h}^*_n$ is calculated based on its similarity with some popular antivaccine-related hashtags. We collected a set of popular antivaccine-related hashtags (please refer to Section~\ref{sec:IC} for details), extracted the feature representations of these hashtags, and then averaged these representations as a single vector, $\overline{{\bm h}}_{\textit{anti}}$, to represent the popular hashtags. To compute the similarity between ${\bm h}^*_n$ and $\overline{{\bm h}}_{\textit{anti}}$, we first passed ${\bm h}^*_n$ through another multilayer perceptron (MLP$_2$) such that the output had the same dimension as ${\overline{{\bm h}}_\textit{anti}}$: 
\begin{equation}
{\bm v}^h_n = tanh({\bm W}_{H_2} {\bm h}^*_n + {\bm b}_{H_2}), \ n \in [1, N]
\end{equation}

We then computed the task-level attention of the hashtag, $p_n^{\textit{Ta}}$, via the similarity between the $n$-th hashtag and the representative antivaccine hashtag $\overline{{\bm h}}_{\textit{anti}}$:
\begin{equation}
p_n^{\textit{Ta}} = {{\bm v}^h_n}^T\overline{\bm h}_{\textit{anti}}, \ n\in[1, N]
\end{equation}

The semantic-level and task-level similarities were combined:
\begin{equation}
p_n^{\textit{SeTa}} = p_n^{\textit{Se}} + p_n^{\textit{Ta}}, n\in[1,N]
\end{equation}
Then, $p_n^{\textit{SeTa}}$ was passed through a softmax function to obtain the final attention weight values $\alpha^{\textit SeTa}_n$ for the $n$-th hashtag. 

We were then able to compute the vector representation of all the hashtags in the caption with their SeTa-attentions:
\begin{equation}
{\bm F}^{\textit{att}}_H = \sum_{n=1}^{N}\alpha^{\textit{SeTa}}_n{\bm h}^*_n, \ n\in[1, N]
\end{equation}

\subsection{The Text Branch}
The text branch primarily processes captions and text that overlays the images. Text overlaid on an image can also provide useful information for detecting antivaccine messages. For example, in Fig.~\ref{fig:r1}(a), the words {\it mandate, force, deceive, or threaten those who question it} are useful in classifying the post as an antivaccine message. We employed the popular Tesseract OCR algorithm\footnote{https://pypi.org/project/pytesseract/} to extract textual content from images. The extracted textual content was concatenated to the caption and then fed into the text branch for further processes, as illustrated in Fig.~\ref{fig:r2}. The texts in the caption or those extracted from OCR consist of sequential natural English words and punctuation. To represent textual contents efficiently, we chose to vectorize texts with the fine-tuned fastText model. Words in the texts are transformed into vector representations by the following embedding matrix:
\begin{equation}
{\bm w}_t = {\bm E}_{\textit{fast}}{\bm o}_t,\  t \in [1, T]
\end{equation}
where ${\bm o}_t$ is the one-hot encoding representation of the $t$-th word in a text sequence with $T$ words; and ${\bm w}_t$ is the corresponding $t$-th word representation in the text. After obtaining the word-level representations, we investigated the sentence-level semantic meaning of each word in the text. A bidirectional RNN with GRU was utilized to catch both the forward and backward context information of each word within the text:
\begin{equation}
\overleftrightarrow{\bm{w}}_t = \left[\overleftarrow{\textbf{biGRU}}({\bm w}_t);\ \overrightarrow{\textbf{biGRU}}({\bm w}_t)\right]
\end{equation}
where $\overrightarrow{\textbf{biGRU}}({\bm w}_t)$ and $\overleftarrow{\textbf{biGRU}}({\bm w}_t)$ correspond to forward and backward hidden context states of the $t$-th word in the text, respectively. After obtaining each word's context representation, we followed the same concept introduced in Section~\ref{sec:33} to compute semantic- and task-level attention weights $\beta^{\textit SeTa}_t$ for the text branch. The whole text representation ${\bm F}^{\textit{att}}_C$ could then be represented as:
\begin{equation}
{\bm F}^{\textit{att}}_C = \sum_{t=1}^{T} \beta^{\textit SeTa}_t\overleftrightarrow{\bm{w}}_t, \ t\in[1, T]
\end{equation}

\subsection{The Image Branch}
\label{sec:ImgBranch}
To extract visual features from images, we used a fine-tuned VGG19 network. The fine-tune details were described in Section~\ref{sec:IC}. In this study, we extracted hidden representations ${\bm Z}_V$ from the last convolutional layer of the fifth convolutional block in the fine-tuned VGG19 network. 
Attention mechanisms have been proven to be successful in both natural language processing~\cite{r13} and computer vision areas~\cite{xu2015}. The visual-based attention mechanism usually helps the model pay attention to the important spatial areas of the input feature maps. In the image branch, we employed the idea of \textit{concurrent spatial and channel squeeze-and-excitation} (scSE) block~\cite{r27,RoyMICCAI18}. The scSE block could help the model pay attention to different areas within feature maps and pay attention to discriminative feature channels. After applying the scSE block on ${\bm Z}_V$ to generate spatial-and-channel attention weights ${\bm s}_V$, we created the weighted image feature maps:
\begin{equation}
{\bm Z}^{\textit{att}}_V = {\bm Z}_V {\bm s}_V
\end{equation}
where ${\bm Z}^{\textit{att}}_V$ is the attention-weighted feature map. We then passed this feature map through a global average pooling (GAP) layer~\cite{r43} to generate the final image feature:
\begin{equation}
{\bm F}^{\textit{att}}_V = \textbf{GAP}({\bm Z}^{\textit{att}}_V)
\end{equation}
The GAP operation reduced feature map dimensions and decreased the total number of parameters in our model.

\subsection{Projection and Fusion Modules}
\subsubsection*{Feature Projection} The multimodal features from three branches were in different feature spaces. Directly fusing them by simple operations, such as summation, was not suitable. We proposed a projection module that contained one fully-connected layer with a rectified linear unit (ReLU) activation function to project the features from three branches to the same representation space. The image branch features after the projection can be computed as follows:
\begin{equation}
{\bm F}^{P}_V = {\textbf P_M}({\bm F}^{\textit{att}}_V)
\end{equation}
where ${\textbf P_M}$ denotes the feature project module. Similarly, we obtained the projected caption features ${\bm F}^{P}_C$ and projected hashtag features ${\bm F}^{P}_H$.

\subsubsection*{Feature Fusion} 
After obtaining the projected multimodal features, we were able to compute the SeTa attentions $\gamma^{SeTa}$ on these features as before. We then combined these weighted features by weighted summation to generate a fused feature ${\bm F}^{att}_{F}$ as follows:
\begin{equation}
{\bm F}^{att}_{F} = \sum_i \gamma_i^{SeTa}{\bm F}^{P}_i,\quad i= [H,C,V]
\end{equation}
multimodal independent features were concatenated with the fused feature to create the following comprehensive feature:
\begin{equation}
{\bm F} = [{\bm F}^{att}_H; {\bm F}^{att}_C; {\bm F}^{att}_V; {\bm F}^{att}_F]
\end{equation}
\begin{figure}[!t]
\centering
\includegraphics[width=\linewidth]{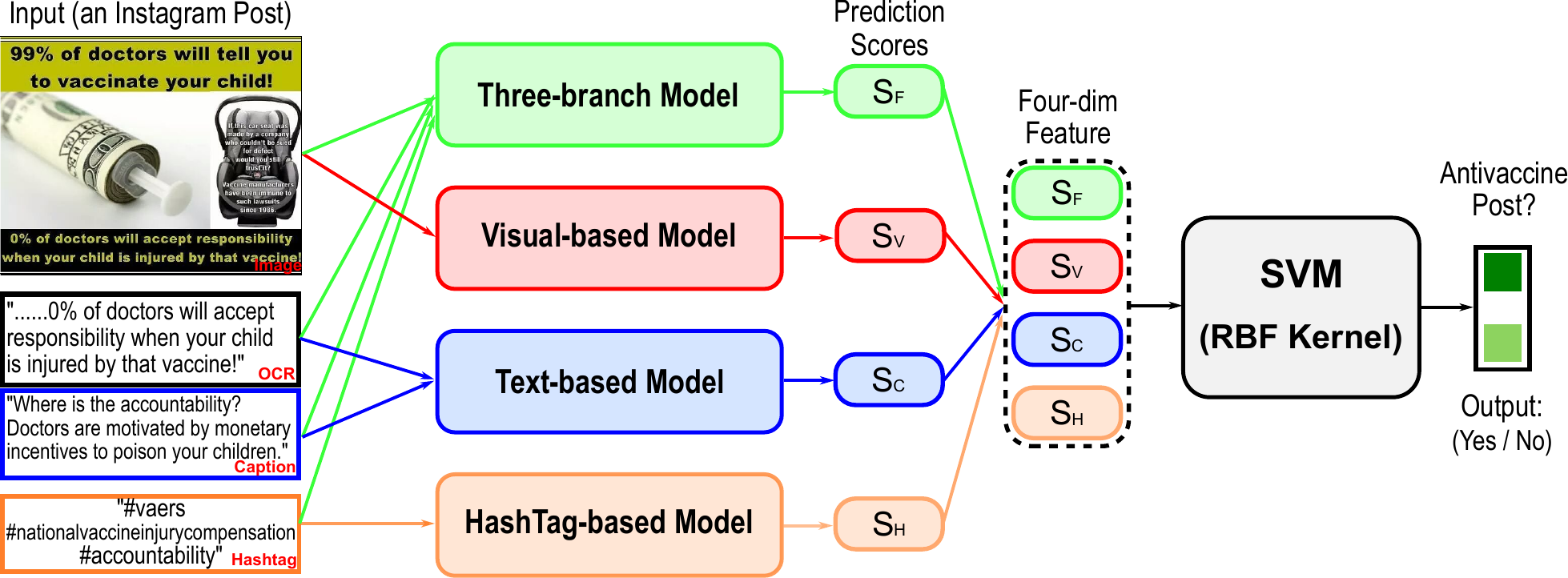}
\caption{Flow chart of the ensemble method for antivaccine message detection. }
\label{fig:r6}
\end{figure}

\subsection{Classifier}
Given the comprehensive feature ${\bm F}$, the probability of this message belonging to an antivaccine message or not, $p(\bm F)$, was able to be predicted by a classifier. In this paper, we built the classifier with three dense layers. During the model training, we used the cross-entropy to compute the loss of the $k$-th input message as follows:
\begin{equation}
\mathcal{L}({\bm F}_k) = -\big[y_k\textit{log}\ (p({\bm F}_k)) + (1-y_k)\textit{log}\ (1 - p({\bm F}_k))\big]
\end{equation}
where $k\in[1, K]$ and $K$ is the total number of input data samples; ${\bm F}_k$ is the comprehensive feature of the $k$-th message; and $y_k$ is the ground truth label of the $k$-th input message.

\subsection{Ensemble of the multimodal Networks}
\label{sec: Ensemble}
Ensemble methods leverage multiple models to achieve better testing performance than any single model alone. Ensemble methods have been successfully applied to different detection and classification tasks~\cite{Cao2020, Kumar2017, Han2014}. For our task, after the proposed multimodal network was trained, it could be applied in order to test the dataset and made predictions directly. To further improve the model performance, an ensemble method was proposed to employ both single-modal prediction results and multimodal network prediction results. Fig.~\ref{fig:r6} illustrates the whole ensemble method process, which combines four different model outcomes to generate final predictions. In the figure, the {\bf three-branch model} denotes the proposed multimodal network in this paper; the {\bf visual-based model} denotes the single modal network using images only; the {\bf text-based model} denotes the single modal network using captions and OCR results only; the {\bf hashtag-based model} denotes the single modal network using hashtags only. We trained these four models with the same training dataset. For each training data, we first used the multimodal network to obtain a prediction score -- $s_F$; secondly, we used the three single-modal networks to make individual predictions and get three prediction scores -- $s_V$, $s_C$, and $s_H$; thirdly, these four prediction scores were combined as a four-dimensional feature: $(s_F, s_V, s_C, s_H)$. For all the training samples, we acquired their corresponding four-dimensional features. Finally, an SVM model with a radial basis function (RBF) kernel was trained with these four-dimensional features. In the testing phase, for all testing samples, four-dimensional features were generated. These new features were then fed into the trained SVM model to make final ensemble predictions.                

\section{Experiment and Results}
In this section, we introduced a real-world social media dataset collected from Instagram and then presented the experimental implementation details. To evaluate the effectiveness of each component in the proposed model, we tested the proposed model on this dataset and compared the performance with several state-of-the-art models. We then performed ablation studies on several variations of the proposed network. Finally, we conducted an online antivaccine message detection experiment with our proposed model. 

\subsection{Dataset}
\label{sec:Data}
We collected Instagram posts posted from January 2016 to October 2019 using an Instagram API toolkit\footnote{https://pypi.org/project/InstagramAPI/}. The total number of collected samples included 31,282 Instagram posts that consisted of 50\% antivaccine posts and 50\% non-antivaccine posts. For the data collection, we adopted a snowballing method widely used for systematic literature reviews~\cite{wohlin2014}. The first step of snowballing involved searching for posts using various combinations of keywords and hashtags that are commonly found in antivaccine posts on social media\cite{kata2012, fortin2011}. Examples of such keywords are {\it \#vaccinesafety}, {\it \#bigpharma}, {\it \#informedconsent}, and {\it \#adverseeffects} \cite{kata2012, fortin2011}. From these search results, we downloaded antivaccine posts. Furthermore, from the search results, we found Instagram accounts that frequently post antivaccine messages and have large numbers of followers. These accounts are representatives of antivaccinists as evidenced by the large numbers of antivaccine messages they post and their large numbers of followers. We found a total of 32 accounts with an average of 8,368 followers per account. These antivaccine accounts include Instagram communities and individual antivaccinists. From these accounts, we downloaded additional antivaccine messages. Non-antivaccine posts included both provaccine posts and vaccine-irrelevant posts and were collected in a similar manner. We first used combinations of known keywords and hashtags, such as {\it \#vaccinesaves} and {\it \#vaccineworks}, to find noticeable provaccine accounts. Provaccine posts and vaccine-irrelevant posts were collected from 29  representative accounts. These provaccine accounts included globally or nationally recognizable institutions, such as the Centers for Disease Control (CDC), the World Health Organization (WHO), and Stanford Medical School, in addition to verified doctors. Their known credibility ensures they represent provaccine communities on social media. We engaged three trained annotators to label these posts independently. They then used a majority voting scheme to choose the ultimate labels for the posts without a consensus among the three annotators. For training the proposed model, we randomly split the whole dataset into the training, validation, and testing sets by a 7:1:2 ratio. In other words, 21,000 samples were used as the training set, 3,000 samples were used as the validation set, and the remaining samples were used as our testing samples.
\begin{table}[!t]
\fontsize{8}{12}\selectfont
\centering
\caption{Results of different models on the Instagram dataset}
\begin{tabular}{ccccccc}
\toprule
No. & Models & Accuracy & Precision & Recall & $F_1$ \\
\midrule
1 & VGG16~\cite{Simonyan15} & 0.827 & 0.809 & 0.829 & 0.819 \\
2 & VGG19~\cite{Simonyan15} & 0.829 & 0.822 & 0.812 & 0.817 \\
3 & ResNet50~\cite{r10} & 0.838 & 0.870 & 0.771 & 0.818 \\
4 & ResNet101~\cite{r10} & 0.839 & 0.847 & 0.803 & 0.825 \\
5 & DenseNet121~\cite{Huang17} & 0.833 & 0.866 & 0.761 & 0.810 \\
6 & DenseNet169~\cite{Huang17} & 0.841 & 0.886 & 0.759 & 0.818 \\
\midrule
7 & GRU2~\cite{MaIJCAI16} & 0.884 & 0.877 & 0.875 & 0.876 \\
8 & GRU1~\cite{MaIJCAI16} & 0.879 & 0.869 & 0.874 & 0.871 \\
9 & LSTM1~\cite{MaIJCAI16} & 0.846 & 0.861 & 0.803 & 0.831 \\
\midrule
10 & att-RNN~\cite{r26} & 0.920 & 0.912 & 0.917 & 0.915 \\
11 & EANN~\cite{r21} & 0.848 & 0.845 & 0.828 & 0.836 \\
12 & MVAE~\cite{KhattarWWW2019} & 0.929 & 0.936 & 0.929 & 0.933 \\
13 & three-branch (ours) & 0.966 & 0.969 & 0.957 & 0.963 \\
14 & ensemble (ours) & \textbf{0.974} & \textbf{0.978} & \textbf{0.967} & \textbf{0.973} \\
\bottomrule
\end{tabular}
\label{tab:r1}
\end{table}

\subsection{Implementation Configurations}
\label{sec:IC}
\subsubsection*{The Hashtag Branch} We fine-tuned the pre-trained fastText word embedding model by continuing the model training with auxiliary data samples. The hashtag branch input size was set as 30 because almost no post contains more than 30 hashtags. Moreover, in the SeTa-attention mechanism, based on their frequency, we chose popular hashtags in antivaccine posts on social media reported in a prior study~\cite{xu2019}. Selected examples are {\it \#vaccinetruth}, {\it \#vaccineinjury}, {\it \#vaccinecauseautism}, {\it \#vaccineawareness}, {\it \#fascism}, {\it \#whistleblower}, {\it \#bigpharma}, and {\it \#informedconsent}.
\subsubsection*{The Text Branch} We used the same embedding representation as in the hashtag branch. The text branch input size was set as 680. It was the maximum word length of both captions and OCR results.
\subsubsection*{The Image Branch} Firstly, we set the fifth convolutional block of the VGG19 network to be trainable and then fine-tuned the pre-trained VGG19 model. The output of the last convolutional layer of the VGG19 model was extracted as the image feature maps with a dimension of $7\times7\times512$. For the scSE block of the image branch, we used the default value of reduction ratio as 16.
\subsubsection*{Classifier and Training} After we obtained the multimodal comprehensive features, they were passed through three dense layers with hidden units of 256, 128, and 64, respectively. Each dense layer was followed by a batch normalization layer and a dropout layer. During model training, we chose an RMSprop optimizer with a learning rate from $1\times10^{-6}$ to $1\times10^{-4}$ with a mini-batch size of 32. Moreover, the proposed model was implemented with Keras, which uses Tensorflow as the backend engine.

\subsection{Quantitative Performance Comparison}
\label{sec:PC}
In this section, we compare the proposed model to several state-of-the-art models on our dataset. Table~\ref{tab:r1} displays the performance of all the compared models in three categories: (1)~image-based networks: these models are listed from No.~1 to No.~6 in the table. In our experiment, these models were initialized with weights learned from ImageNet and fine-tuned with our dataset; (2)~text-based networks: these models are listed from No.~7 to No.~9 in the table. These three networks were designed based on a recurrent neural network with GRU or long short-term memory (LSTM) units. They were trained from scratch on our dataset; and (3)~multimodal based networks: these models are listed from No.~10 to No.~12 in the table, which uses both visual and textual information for the model training. These models were also trained entirely on our dataset. The performances of our three-branch multimodal network and its ensemble version are listed in the last two lines of the table. 

Based on the testing results in Table~\ref{tab:r1}, we have several observations: (1)~our proposed network outperforms the other multimodal based networks; (2)~models that leverage multimodal information perform better than models only considering single modal information. It also validates the multimodal information is important to detect antivaccine posts on social media; (3)~text-based models perform better than image-based models. This is reasonable since textual information contains more explicit antivaccine information than visual information in the posts; and (4)~the proposed ensemble method improves the model performance. 

\subsection{Ablation Study}
To evaluate the effectiveness of main components in our multimodal network ({\bf three-branch}), we conducted more experiments to examine their corresponding performances: (1)~{\bf ours\_noF}: this was designed as the proposed multimodal network but without using the fused feature ${\bm F}^{att}_F$; (2)~{\bf ours\_noP}: this was created as the proposed network without the projection module (i.e., the three multimodal features were leveraged directly); (3)~{\bf ours\_noAtt}: all the attention mechanism parts of the proposed network were removed in this variation; (4)~{\bf ours\_noOCR}: this was created as the proposed multimodal network without the OCR component; (5)~{\bf single-modal}: we create three single-modal networks. Each of them is one branch of the proposed model. For example, \textit{image\_only} refers to a model that consists of the image branch of the proposed model only; (6)~{\bf bi-modal}: each bi-modal network takes only two branches of the proposed network. For instance, \textit{image+caption} denotes the model that has both image and text branches of the proposed network but does not include the hashtag branch.

Based on the results in Table~\ref{tab:r2}, we observe the following: (1)~based on the results of No.~0 and No.~1, the fused feature ${\bm F}^{att}_F$ has positive effects for antivaccine detection; (2)~based on the results of No.~0 and No.~2, the model performance is decreased if there is no projection module; (3)~based on the results of No.~0 and No.~3, the proposed model performance benefits from the attention mechanisms. It helps the proposed model improve the testing accuracy from 94.2\% to 96.6\%. It demonstrates the importance of the proposed attention mechanisms in the model; (4)~based on the results of No.~0 and No.~4, the OCR module can improve the model performance slightly because not every post contains text overlaid on the image; (5)~results from No.~5a to No.~5c illustrate the single caption branch has better testing performance than the single image branch. The results also demonstrate that the textual information plays a more important role than visual information for our antivaccine detection task; and (6)~results from No.~6a to No.~6c indicate that bi-modal networks have better performances than single-modal networks, but bi-modal networks still have the inferior performances compared to our multimodal network that uses all three modalities. It also demonstrates that it is important to leverage all the multimodal information to detect antivaccine posts on social media.  
\begin{table}[!t]
\fontsize{8}{12}\selectfont
\centering
\caption{Ablation study on the proposed network} 
\begin{tabular}{cccccc}
\toprule
No. & Models & Accuracy & Precision & Recall & $F_1$ \\
\midrule
0 & three-branch (ours) & 0.966 & 0.969 & 0.957 & 0.963 \\
1 & ours\_noF & 0.953 & 0.968 & 0.931 & 0.949 \\
2 & ours\_noP & 0.951 & 0.967 & 0.928 & 0.947 \\
3 & ours\_noAtt & 0.942 & 0.955 & 0.919 & 0.937 \\
4 & ours\_noOCR & 0.964 & 0.975 & 0.948 & 0.961 \\
\midrule
5a & image\_only & 0.865 & 0.886 & 0.817 & 0.850 \\
5b & caption\_only & 0.911 & 0.908 & 0.902 & 0.905 \\
5c & tag\_only & 0.855 & 0.953 & 0.728 & 0.826 \\
\midrule
6a & image+caption & 0.934 & 0.944 & 0.913 & 0.929 \\
6b & image+tag & 0.926 & 0.946 & 0.895 & 0.920 \\
6c & caption+tag & 0.927 & 0.934 & 0.910 & 0.921 \\
\midrule
7 & ensemble\_mean & 0.972 & 0.974 & 0.967 & 0.970 \\
8 & ensemble\_max  & 0.970 & 0.972 & 0.963 & 0.967 \\
9 & ensemble\_vote & 0.971 & 0.974 & 0.965 & 0.969 \\
10 & ensemble (ours) & 0.974 & 0.978 & 0.967 & 0.973 \\
\bottomrule
\end{tabular}
\label{tab:r2}
\end{table}

For the proposed ensemble method, given the four prediction scores (i.e., ${s_F}$, ${s_V}$, ${s_C}$, and ${s_H}$, as described in Section~\ref{sec: Ensemble}), the range of each prediction score is $[0, 1]$. $0$ denotes non-antivaccine posts, and $1$ denotes antivaccine posts. To evaluate the effectiveness of the proposed ensemble method ({\bf ensemble}), we conducted several additional experiments to compare the performances: (7)~{\bf ensemble\_mean}:  the mean value of these four scores were computed as the final prediction score; (8)~{\bf ensemble\_max}: we first computed the absolute differences between prediction scores and $0.5$, then we chose the prediction score that had the maximum absolute difference value as the final prediction score. For example, suppose the four prediction scores for an input post are $0.4$, $0.3$, $0.35$, and $0.9$, respectively. The absolute values after subtracting $0.5$ are $0.1$, $0.2$, $0.15$, and $0.4$. We can thus consider this post as an antivaccine post because its prediction is $0.9$, which has the maximum absolute difference value (i.e., $0.4$) in this example; (9)~{\bf ensemble\_vote}: firstly, we made the predictions based on the four scores individually and then made our final prediction with the majority voting rule. If a tie occurred, we took the mean values of both sides and then chose the one with a maximum deviation from $0.5$ as the final prediction result. For example, if the prediction scores are $0.2$, $0.3$, $0.8$, and $0.9$, then the corresponding prediction results are $0$, $0$, $1$, and $1$, wherein a tie occurs. In this instance, we obtained the mean values of both sides: $0.25$ and $0.85$. We then acquired the values of deviation from $0.5$ (i.e., the absolute difference values from $0.5$) as $0.25$ and $0.35$. We could thus consider this example as an antivaccine post as $0.35$ is the maximum deviation value from $0.5$ in this example, and the corresponding prediction results are both $1$ (i.e., antivaccine). 

Base on the results No.~7 to No.~10 in Table~\ref{tab:r2}, we discovered that the proposed ensemble method achieves a better performance than the other three methods. 
\begin{figure}[!t]
\centering
\includegraphics[width=\linewidth]{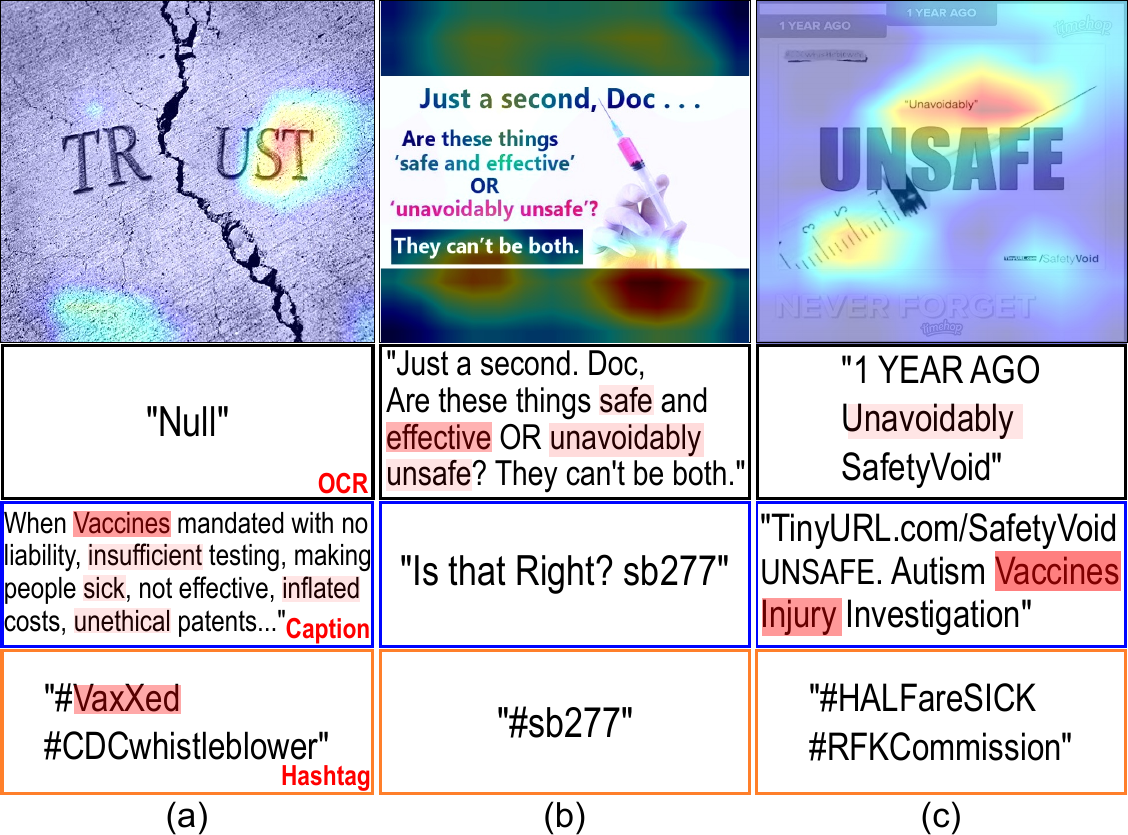}
\caption{Examples of antivaccine posts that are detected successfully by the proposed multimodal network but missed by single-modal networks. Both visual and textual attention mechanism results are illustrated. Red denotes attention weights for both images and texts. The OCR results are illustrated inside black boxes. ``Null'' denotes that no textual content is detected in the post.}
\label{fig:r8}
\end{figure}
\subsection{Qualitative Examples}
To illustrate the importance of the multimodal feature fusion to antivaccine message detection, we ran several experiments with the multimodal network and single-modal networks. We analyzed some experimental results that were detected successfully by the proposed multimodal network but missed by single-modal networks. Furthermore, the visualizations of attention mechanisms provided some straightforward explanations for what the proposed model pays attention to when it makes prediction results. 

As indicated in Fig.~\ref{fig:r8}(a), we present an example of a post that is detected by the proposed multimodal network but missed by the single image model. The image contains little antivaccine information. However, the caption and hashtags provide deterministic clues to identify the category of this post. Fig.~\ref{fig:r8}(b) displays an example of an antivaccine post that is detected by our multimodal network but missed by the single caption model. The image attached in the post evidently presents an attitude of vaccine suspicion. The OCR results contribute when the model makes a prediction. However, the caption does not state a clear attitude toward vaccination. In Fig.~\ref{fig:r8}(c), an antivaccine post is detected by the multimodal network but missed by the single hashtag model. The post contains two hashtags that are not directly related to antivaccine messages; it thus cannot be recognized by the single hashtag model. However, the proposed multimodal network can obtain reliable clues to make an accurate prediction based on the contents of the image and the caption. Based on these examples, we can also observe that the proposed model, which leverages both visual and textual information, outperforms single-modal networks. 
\begin{figure}[!t]
\centering
\includegraphics[width=\linewidth]{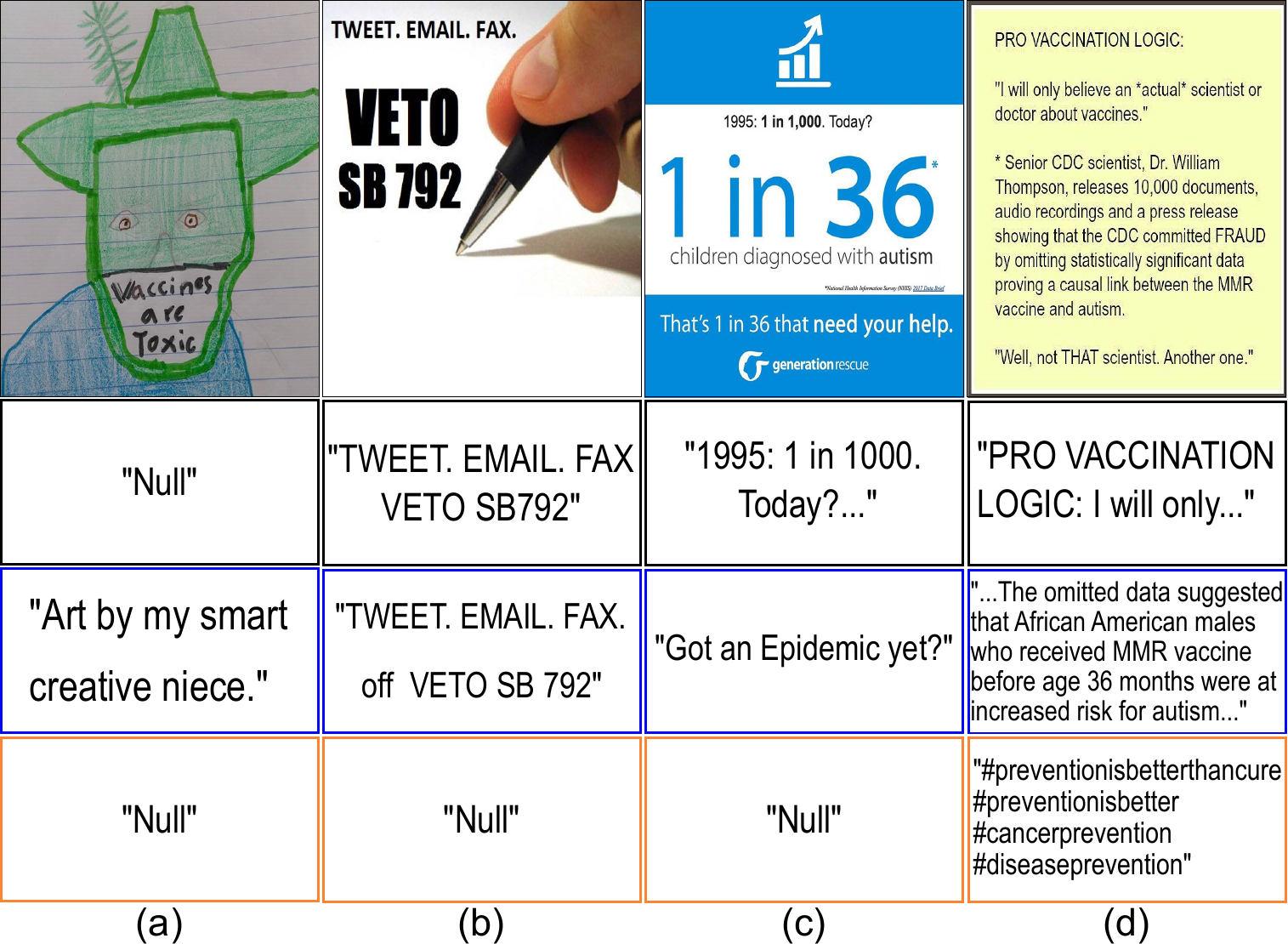}
\caption{Examples of testing failure cases when applying the proposed multimodal network. ``Null'' denotes no hashtags in the post.}
\label{fig:r7}
\end{figure}

\subsection{Failure Case Analysis}
As demonstrated in Table~\ref{tab:r1}, the proposed model achieved 97.3\% F$_1$ score on the Instagram dataset collected from January 2016 to October 2019. This is close to a perfect performance, and it is worth analyzing the failure cases for the future work. Some antivaccine posts contain insufficient information in the images and captions, which are difficult for the proposed model to make correct predictions. As indicated in Fig.~\ref{fig:r7}(a), the post does not contain enough information; only the words on the white mask are relevant. However, the OCR algorithm cannot detect these small, handwritten words correctly. Thus, the proposed model fails to recognize it as an antivaccine post. For some posts, external knowledge is necessary to make correct predictions. For example, in Fig.~\ref{fig:r7}(b), the post is related to rejecting SB792, which concerns child immunization requirements in childcare centers\footnote{http://www.sdiz.org/documents/Sch-CC/sb792\_factsheet\_adapted.pdf}. It is difficult for the proposed model to make correct decisions without the necessary domain knowledge. Moreover, some posts do not contain explicit antivaccine information, as shown in Fig.~\ref{fig:r7}(c). In the picture, only the number of children with autism are illustrated. It is not easy for the model to make correct predictions based only on this post. However, if we read the post exhibited in Fig.~\ref{fig:r7}(d), we understand that antivaccinists believe that there is a relation between vaccines and autism. If the model can infer the context relations among posts and be embedded with the domain knowledge from human experts, this will improve the model performance.

\subsection{Online Antivaccine Message Detection}
\begin{figure}[!t]
\centering
\includegraphics[width=\linewidth]{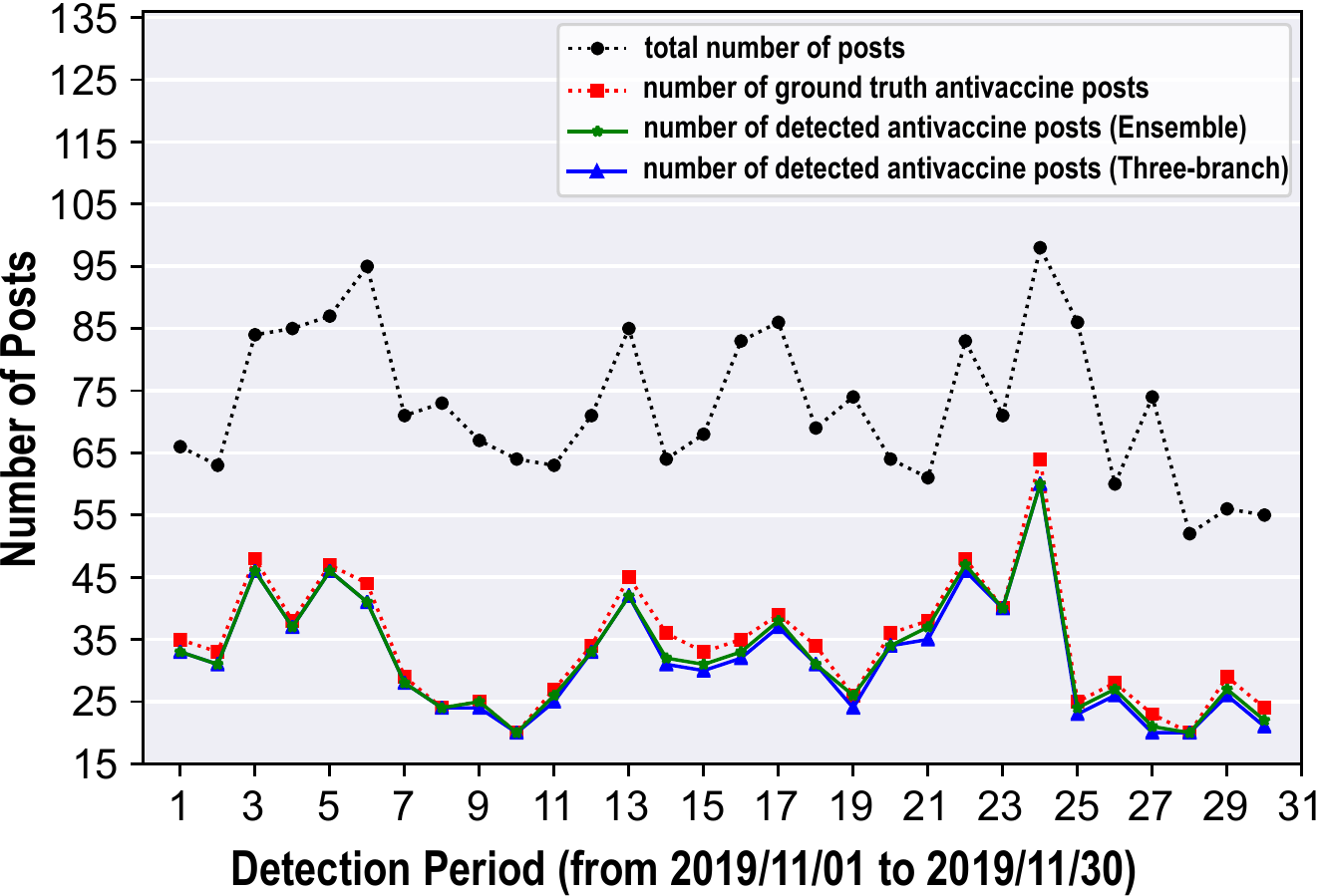}
\caption{Detecting antivaccine-related Instagram posts for 30 days with the proposed multimodal network.}
\label{fig:r5}
\end{figure}

To evaluate the antivaccine message detection ability of the proposed network, we applied the proposed multimodal network on Instagram to detect antivaccine posts for one month (2019/11/01--2019/11/30). The same hashtags that were reported in Section~\ref{sec:Data} were used to search for daily posts that are related to vaccine, antivaccine, or provaccine messages. We then applied the proposed multimodal network to detect the number of antivaccine posts every day. In the practical application, we are concerned about whether the system can identify all antivaccine messages spreading online. In this real-world detection scenario, we reported the number of detected antivaccine messages every day and compared the results with ground truth. Due to the restriction of the Instagram API toolkit, we were only permitted to download less than 80 recent posts for a single hashtag. Once we had obtained the posts, some of them were removed, such as posts not written in English. The total number of vaccine-related posts downloaded within the 30 days were 2,174, of which 1,027 were antivaccine posts. The online detection results are indicated in Fig.~\ref{fig:r5}. Our multimodal network obtained the average precision and recall values of $0.951$ and $0.940$, respectively. Moreover, the proposed ensemble method achieved $0.961$ and $0.958$ for the average precision and recall values separately. The performances are close to the values of No.~13 and No.~14 in Table~\ref{tab:r1}. Therefore, it is feasible that we can apply the proposed multimodal network for antivaccine message detection to social media communities, such as Instagram. Eventually, the detection can assist us in identifying key persuasion tactics to reduce vaccine hesitancy.

\section{Conclusion and Future Work}
In this paper, we studied antivaccine content detection on social media. To detect antivaccine messages, we first constructed a real-world antivaccine message dataset. We then proposed a deep learning neural network that leverages multimodal features. We also proposed a new way to compute the semantic- and task-level attention weight to help the model learn better multimodal feature representations. Moreover, to improve the model performance, we designed an ensemble method to leverage various model outputs to generate final prediction results. Our experimental results exhibited that the proposed network outperformed other relevant models in antivaccine post detection. Furthermore, the proposed model was applied to detecting antivaccine messages online for one consecutive month. The detection results demonstrate the feasibility of the proposed network for antivaccine misinformation detection on social media. 

However, the proposed model still has some limitations when the model is applied to antivaccine posts without sufficient context information. Moreover, the model cannot make correct predictions when human domain knowledge is necessary. For future works, we will consider adding memory cells to force the model to remember some essential features during the model training phase. The memory cells can help the model build connections among different training samples and correctly make predictions for difficult testing cases.

\section{Acknowledgment}
Zuhui Wang and Zhaozheng Yin were supported by NSF grants IIS-2019967, CMMI-1646162, and SUNY Empire Innovation Program (EIP). Young Anna Argyris was supported by Michigan State University's Science + Society @ State, and Trifecta Initiative Facilitating Funds Awards.

\bibliographystyle{IEEEtran}
\bibliography{BHIbibfile}

\begin{thebibliography}{10}
\providecommand{\url}[1]{#1}
\csname url@samestyle\endcsname
\providecommand{\newblock}{\relax}
\providecommand{\bibinfo}[2]{#2}
\providecommand{\BIBentrySTDinterwordspacing}{\spaceskip=0pt\relax}
\providecommand{\BIBentryALTinterwordstretchfactor}{4}
\providecommand{\BIBentryALTinterwordspacing}{\spaceskip=\fontdimen2\font plus
\BIBentryALTinterwordstretchfactor\fontdimen3\font minus
  \fontdimen4\font\relax}
\providecommand{\BIBforeignlanguage}[2]{{%
\expandafter\ifx\csname l@#1\endcsname\relax
\typeout{** WARNING: IEEEtran.bst: No hyphenation pattern has been}%
\typeout{** loaded for the language `#1'. Using the pattern for}%
\typeout{** the default language instead.}%
\else
\language=\csname l@#1\endcsname
\fi
#2}}
\providecommand{\BIBdecl}{\relax}
\BIBdecl

\bibitem{r1}
R.~M. Carpiano and N.~S. Fitz, ``Public attitudes toward child
  undervaccination: A randomized experiment on evaluations, stigmatizing
  orientations, and support for policies,'' \emph{Social Science \& Medicine},
  vol. 185, pp. 127--136, 2017.

\bibitem{r37}
T.~Burki, ``Vaccine misinformation and social media,'' \emph{The Lancet Digital
  Health}, vol.~1, no.~6, pp. e258--e259, 2019.

\bibitem{paige2017}
S.~R. Paige, J.~L. Krieger, and M.~L. Stellefson, ``The influence of ehealth
  literacy on perceived trust in online health communication channels and
  sources,'' \emph{Journal of health communication}, vol.~22, no.~1, pp.
  53--65, 2017.

\bibitem{Perrin2019}
\BIBentryALTinterwordspacing
A.~Perrin and M.~Anderson. Share of u.s. adults using social media, including
  facebook, is mostly unchanged since 2018. [Online]. Available:
  \url{https://www.pewresearch.org/fact-tank/2019/04/10/
  share-of-u-s-adults-using-social-media-including-facebook-is-mostly-unchanged-since-2018/}
\BIBentrySTDinterwordspacing

\bibitem{Eftekhar14}
A.~Eftekhar, C.~Fullwood, and N.~Morris, ``Capturing personality from facebook
  photos and photo-related activities: How much exposure do you need?''
  \emph{Computers in Human Behavior}, vol.~37, pp. 162--170, 2014.

\bibitem{Chancellor16}
S.~Chancellor, J.~A. Pater, T.~A. Clear, E.~Gilbert, and M.~D. Choudhury,
  ``{\#}thyghgapp: Instagram content moderation and lexical variation in
  pro-eating disorder communities,'' in \emph{Proceedings of the 19th {ACM}
  Conference on Computer-Supported Cooperative Work {\&} Social Computing,
  {CSCW} 2016, San Francisco, CA, USA, February 27 - March 2, 2016}.\hskip 1em
  plus 0.5em minus 0.4em\relax {ACM}, 2016, pp. 1199--1211.

\bibitem{Laat2017}
P.~B. de~Laat, ``Big data and algorithmic decision-making: can transparency
  restore accountability?'' \emph{ACM SIGCAS Computers and Society}, vol.~47,
  no.~3, pp. 39--53, 2017.

\bibitem{synnott2017}
J.~Synnott, A.~Coulias, and M.~Ioannou, ``Online trolling: the case of
  madeleine mccann,'' \emph{Computers in Human Behavior}, vol.~71, pp. 70--78,
  2017.

\bibitem{MitraICWSM2016}
T.~Mitra, S.~Counts, and J.~W. Pennebaker, ``Understanding anti-vaccination
  attitudes in social media,'' in \emph{Proceedings of the 10th International
  Conference on Web and Social Media}.\hskip 1em plus 0.5em minus 0.4em\relax
  {AAAI} Press, 2016, pp. 269--278.

\bibitem{BjarkeCoRR2019}
B.~M{\o}nsted and S.~Lehmann, ``Algorithmic detection and analysis of
  vaccine-denialist sentiment clusters in social networks,'' \emph{CoRR}, vol.
  abs/1905.12908, 2019.

\bibitem{r7}
T.~S. Tomeny, C.~J. Vargo, and S.~El-Toukhy, ``Geographic and demographic
  correlates of autism-related anti-vaccine beliefs on twitter, 2009-15,''
  \emph{Social Science \& Medicine}, vol. 191, pp. 168--175, 2017.

\bibitem{r26}
Z.~Jin, J.~Cao, H.~Guo, Y.~Zhang, and J.~Luo, ``Multimodal fusion with
  recurrent neural networks for rumor detection on microblogs,'' in
  \emph{Proceedings of the 25th ACM International Conference on Multimedia},
  no.~22.\hskip 1em plus 0.5em minus 0.4em\relax ACM, 2017, pp. 795--816.

\bibitem{ZhangACMMM2019}
H.~Zhang, Q.~Fang, S.~Qian, and C.~Xu, ``Multi-modal knowledge-aware event
  memory network for social media rumor detection,'' in \emph{Proceedings of
  the 27th {ACM} International Conference on Multimedia}, 2019, pp. 1942--1951.

\bibitem{r21}
Y.~Wang, F.~Ma, Z.~Jin, Y.~Yuan, G.~Xun, K.~Jha, L.~Su, and J.~Gao, ``{EANN:}
  event adversarial neural networks for multi-modal fake news detection,'' in
  \emph{Proceedings of the 24th {ACM} {SIGKDD} International Conference on
  Knowledge Discovery {\&} Data Mining, {KDD}}.\hskip 1em plus 0.5em minus
  0.4em\relax ACM, 2018, pp. 849--857.

\bibitem{KhattarWWW2019}
D.~Khattar, J.~S. Goud, M.~Gupta, and V.~Varma, ``Mvae: Multimodal variational
  autoencoder for fake news detection,'' in \emph{The World Wide Web
  Conference}.\hskip 1em plus 0.5em minus 0.4em\relax ACM, 2019, pp.
  2915--2921.

\bibitem{Schmidt2017}
A.~Schmidt and M.~Wiegand, ``A survey on hate speech detection using natural
  language processing,'' in \emph{Proceedings of the Fifth International
  workshop on natural language processing for social media}.\hskip 1em plus
  0.5em minus 0.4em\relax Association for Computational Linguistics, 2017, pp.
  1--10.

\bibitem{Gasparini2018}
F.~Gasparini, I.~Erba, E.~Fersini, and S.~Corchs, ``Multimodal classification
  of sexist advertisements,'' in \emph{{ICETE} {(1)}}.\hskip 1em plus 0.5em
  minus 0.4em\relax SciTePress, 2018, pp. 565--572.

\bibitem{Krizhevsky2012}
A.~Krizhevsky, I.~Sutskever, and G.~E. Hinton, ``Imagenet classification with
  deep convolutional neural networks,'' in \emph{Advances in neural information
  processing systems}, 2012, pp. 1106--1114.

\bibitem{Mikolov2013}
T.~Mikolov, I.~Sutskever, K.~Chen, G.~S. Corrado, and J.~Dean, ``Distributed
  representations of words and phrases and their compositionality,'' in
  \emph{Advances in neural information processing systems}, 2013, pp.
  3111--3119.

\bibitem{Sabat2019}
B.~O. Sabat, C.~Canton{-}Ferrer, and X.~Gir{\'{o}}{-}i{-}Nieto, ``Hate speech
  in pixels: Detection of offensive memes towards automatic moderation,''
  \emph{CoRR}, vol. abs/1910.02334, 2019.

\bibitem{Simonyan15}
K.~Simonyan and A.~Zisserman, ``Very deep convolutional networks for
  large-scale image recognition,'' in \emph{3rd International Conference on
  Learning Representations}, 2015.

\bibitem{Devlin19}
J.~Devlin, M.~Chang, K.~Lee, and K.~Toutanova, ``{BERT:} pre-training of deep
  bidirectional transformers for language understanding,'' in \emph{{NAACL-HLT}
  {(1)}}.\hskip 1em plus 0.5em minus 0.4em\relax Association for Computational
  Linguistics, 2019, pp. 4171--4186.

\bibitem{kumari20}
K.~Kumari and J.~P. Singh, ``Identification of cyberbullying on multi-modal
  social media posts using genetic algorithm,'' \emph{Transactions on Emerging
  Telecommunications Technologies}, p. e3907, 2020.

\bibitem{r10}
K.~He, X.~Zhang, S.~Ren, and J.~Sun, ``Deep residual learning for image
  recognition,'' in \emph{2016 {IEEE} Conference on Computer Vision and Pattern
  Recognition, {CVPR}}.\hskip 1em plus 0.5em minus 0.4em\relax IEEE Computer
  Society, 2016, pp. 770--778.

\bibitem{yang2019}
F.~Yang, X.~Peng, G.~Ghosh, R.~Shilon, H.~Ma, E.~Moore, and G.~Predovic,
  ``Exploring deep multimodal fusion of text and photo for hate speech
  classification,'' in \emph{Proceedings of the Third Workshop on Abusive
  Language Online}, 2019, pp. 11--18.

\bibitem{Szegedy16}
C.~Szegedy, V.~Vanhoucke, S.~Ioffe, J.~Shlens, and Z.~Wojna, ``Rethinking the
  inception architecture for computer vision,'' in \emph{Proceedings of the
  IEEE conference on computer vision and pattern recognition}.\hskip 1em plus
  0.5em minus 0.4em\relax {IEEE} Computer Society, 2016, pp. 2818--2826.

\bibitem{Gomez20}
R.~Gomez, J.~Gibert, L.~G{\'{o}}mez, and D.~Karatzas, ``Exploring hate speech
  detection in multimodal publications,'' in \emph{The IEEE Winter Conference
  on Applications of Computer Vision}.\hskip 1em plus 0.5em minus 0.4em\relax
  {IEEE}, 2020, pp. 1459--1467.

\bibitem{Choudhury14}
M.~D. Choudhury, M.~R. Morris, and R.~W. White, ``Seeking and sharing health
  information online: comparing search engines and social media,'' in
  \emph{Proceedings of the SIGCHI Conference on Human Factors in Computing
  Systems}.\hskip 1em plus 0.5em minus 0.4em\relax {ACM}, 2014, pp. 1365--1376.

\bibitem{eysenbach08}
G.~Eysenbach, \emph{Credibility of health information and digital media: New
  perspectives and implications for youth}.\hskip 1em plus 0.5em minus
  0.4em\relax MacArthur Foundation Digital Media and Learning Initiative, 2008.

\bibitem{Viviani17}
M.~Viviani and G.~Pasi, ``Credibility in social media: opinions, news, and
  health information - a survey,'' \emph{Wiley Interdisciplinary Reviews: Data
  Mining and Knowledge Discovery}, vol.~7, no.~5, 2017.

\bibitem{Abbasi13}
A.~Abbasi, T.~Fu, D.~Zeng, and D.~A. Adjeroh, ``Crawling credible online
  medical sentiments for social intelligence,'' in \emph{2013 International
  Conference on Social Computing}.\hskip 1em plus 0.5em minus 0.4em\relax
  {IEEE} Computer Society, 2013, pp. 254--263.

\bibitem{Weitzel14}
L.~Weitzel, J.~P.~M. de~Oliveira, and P.~Quaresma, ``Measuring the reputation
  in user-generated-content systems based on health information,'' in
  \emph{2014 International Conference on Computational Science}, vol.~29.\hskip
  1em plus 0.5em minus 0.4em\relax Elsevier, 2014, pp. 364--378.

\bibitem{Ghenai17}
A.~Ghenai and Y.~Mejova, ``Catching zika fever: Application of crowdsourcing
  and machine learning for tracking health misinformation on twitter,'' in
  \emph{2017 IEEE International Conference on Healthcare Informatics}.\hskip
  1em plus 0.5em minus 0.4em\relax {IEEE} Computer Society, 2017, p. 518.

\bibitem{Liu19}
Y.~Liu, K.~Yu, X.~Wu, L.~Qing, and Y.~Peng, ``Analysis and detection of
  health-related misinformation on chinese social media,'' \emph{{IEEE}
  Access}, vol.~7, pp. 154\,480--154\,489, 2019.

\bibitem{friedman2001}
J.~H. Friedman, ``Greedy function approximation: a gradient boosting machine,''
  \emph{Annals of statistics}, pp. 1189--1232, 2001.

\bibitem{botsis2014}
T.~{Botsis}, J.~{Scott}, E.~J. {Woo}, and R.~{Ball}, ``Identifying similar
  cases in document networks using cross-reference structures,'' \emph{IEEE
  Journal of Biomedical and Health Informatics}, vol.~19, no.~6, pp.
  1906--1917, 2015.

\bibitem{CortesV95}
C.~Cortes and V.~Vapnik, ``Support-vector networks,'' \emph{Mach. Learn.},
  vol.~20, no.~3, pp. 273--297, 1995.

\bibitem{Shi2018}
J.~{Shi}, X.~{Zheng}, Y.~{Li}, Q.~{Zhang}, and S.~{Ying}, ``Multimodal
  neuroimaging feature learning with multimodal stacked deep polynomial
  networks for diagnosis of alzheimer's disease,'' \emph{IEEE Journal of
  Biomedical and Health Informatics}, vol.~22, no.~1, pp. 173--183, 2018.

\bibitem{Correa2019}
J.~C. {Vásquez-Correa}, T.~{Arias-Vergara}, J.~R. {Orozco-Arroyave},
  B.~{Eskofier}, J.~{Klucken}, and E.~{Nöth}, ``Multimodal assessment of
  parkinson's disease: A deep learning approach,'' \emph{IEEE Journal of
  Biomedical and Health Informatics}, vol.~23, no.~4, pp. 1618--1630, 2019.

\bibitem{Kawahara2019}
J.~{Kawahara}, S.~{Daneshvar}, G.~{Argenziano}, and G.~{Hamarneh},
  ``Seven-point checklist and skin lesion classification using multitask
  multimodal neural nets,'' \emph{IEEE Journal of Biomedical and Health
  Informatics}, vol.~23, no.~2, pp. 538--546, 2019.

\bibitem{song2018}
J.~Song, K.~Han, D.~Lee, and S.-W. Kim, ````is a picture really worth a
  thousand words?'': A case study on classifying user attributes on
  instagram,'' \emph{PloS One}, vol.~13, no.~10, 2018.

\bibitem{r28}
P.~Bojanowski, E.~Grave, A.~Joulin, and T.~Mikolov, ``Enriching word vectors
  with subword information,'' \emph{Trans. Assoc. Comput. Linguistics}, vol.~5,
  pp. 135--146, 2017.

\bibitem{r13}
Z.~Yang, D.~Yang, C.~Dyer, X.~He, A.~Smola, and E.~Hovy, ``Hierarchical
  attention networks for document classification,'' in \emph{{NAACL} {HLT}
  2016, The 2016 Conference of the North American Chapter of the Association
  for Computational Linguistics: Human Language Technologies}.\hskip 1em plus
  0.5em minus 0.4em\relax ACL, 2016, pp. 1480--1489.

\bibitem{xu2015}
K.~Xu, J.~Ba, R.~Kiros, K.~Cho, A.~C. Courville, R.~Salakhutdinov, R.~S. Zemel,
  and Y.~Bengio, ``Show, attend and tell: Neural image caption generation with
  visual attention,'' in \emph{International Conference on Machine Learning},
  ser. {JMLR} Workshop and Conference Proceedings, vol.~37, 2015, pp.
  2048--2057.

\bibitem{r27}
J.~Hu, L.~Shen, and G.~Sun, ``Squeeze-and-excitation networks,'' in \emph{2018
  {IEEE} Conference on Computer Vision and Pattern Recognition, {CVPR}}.\hskip
  1em plus 0.5em minus 0.4em\relax {IEEE} Computer Society, 2018, pp.
  7132--7141.

\bibitem{RoyMICCAI18}
A.~G. Roy, N.~Navab, and C.~Wachinger, ``Concurrent spatial and channel
  'squeeze {\&} excitation' in fully convolutional networks,'' in \emph{Medical
  Image Computing and Computer Assisted Intervention - {MICCAI}}, 2018, pp.
  421--429.

\bibitem{r43}
\BIBentryALTinterwordspacing
M.~Lin, Q.~Chen, and S.~Yan, ``Network in network,'' in \emph{2nd International
  Conference on Learning Representations, {ICLR} 2014, Banff, AB, Canada, April
  14-16, 2014, Conference Track Proceedings}, Y.~Bengio and Y.~LeCun, Eds.,
  2014. [Online]. Available: \url{http://arxiv.org/abs/1312.4400}
\BIBentrySTDinterwordspacing

\bibitem{Cao2020}
H.~{Cao}, H.~{Liu}, E.~{Song}, G.~{Ma}, R.~{Jin}, X.~{Xu}, T.~{Liu}, and
  C.~{Hung}, ``A two-stage convolutional neural networks for lung nodule
  detection,'' \emph{IEEE Journal of Biomedical and Health Informatics}, pp.
  1--1, 2020.

\bibitem{Kumar2017}
A.~{Kumar}, J.~{Kim}, D.~{Lyndon}, M.~{Fulham}, and D.~{Feng}, ``An ensemble of
  fine-tuned convolutional neural networks for medical image classification,''
  \emph{IEEE Journal of Biomedical and Health Informatics}, vol.~21, no.~1, pp.
  31--40, 2017.

\bibitem{Han2014}
L.~{Han}, S.~{Luo}, J.~{Yu}, L.~{Pan}, and S.~{Chen}, ``Rule extraction from
  support vector machines using ensemble learning approach: An application for
  diagnosis of diabetes,'' \emph{IEEE Journal of Biomedical and Health
  Informatics}, vol.~19, no.~2, pp. 728--734, 2015.

\bibitem{wohlin2014}
C.~Wohlin, ``Guidelines for snowballing in systematic literature studies and a
  replication in software engineering,'' in \emph{Proceedings of the 18th
  International Conference on Evaluation and Assessment in Software
  Engineering}, 2014, pp. 1--10.

\bibitem{kata2012}
A.~Kata, ``Anti-vaccine activists, web 2.0, and the postmodern paradigm--an
  overview of tactics and tropes used online by the anti-vaccination
  movement,'' \emph{Vaccine}, vol.~30, no.~25, pp. 3778--3789, 2012.

\bibitem{fortin2011}
D.~Fortin, M.~Uncles, M.~S. Lee, and M.~Male, ``Against medical advice: the
  anti-consumption of vaccines,'' \emph{Journal of Consumer Marketing}, 2011.

\bibitem{Huang17}
G.~Huang, Z.~Liu, L.~van~der Maaten, and K.~Q. Weinberger, ``Densely connected
  convolutional networks,'' in \emph{2017 {IEEE} Conference on Computer Vision
  and Pattern Recognition, {CVPR} 2017, Honolulu, HI, USA, July 21-26, 2017},
  2017, pp. 2261--2269.

\bibitem{MaIJCAI16}
J.~Ma, W.~Gao, P.~Mitra, S.~Kwon, B.~J. Jansen, K.-F. Wong, and M.~Cha,
  ``Detecting rumors from microblogs with recurrent neural networks,'' in
  \emph{Proceedings of the Twenty-Fifth International Joint Conference on
  Artificial Intelligence}, ser. IJCAI'16.\hskip 1em plus 0.5em minus
  0.4em\relax AAAI Press, 2016, pp. 3818--3824.

\bibitem{xu2019}
Z.~Xu, ``Personal stories matter: topic evolution and popularity among pro-and
  anti-vaccine online articles,'' \emph{Journal of Computational Social
  Science}, vol.~2, no.~2, pp. 207--220, 2019.

\end{thebibliography}

\end{document}